\documentclass[%
 reprint,
 amsmath,amssymb,
 aps,twocolumn
]{revtex4}

\usepackage{natbib}
\usepackage{aas_macros}
\usepackage{graphicx}
\usepackage{dcolumn}
\usepackage{bm}
\usepackage{xcolor}
\usepackage[normalem]{ulem}
\usepackage{verbatim}
\usepackage{enumitem}
\usepackage{orcidlink}
\usepackage{xcolor}

\begin{document}

\title{Topological Signatures of Heating and Dark Matter in the 21 cm Forest}

\author{Hayato Shimabukuro\orcidlink{0000-0003-4850-0656}}
\affiliation{South-Western Institute for Astronomy Research (SWIFAR), Yunnan University, Kunming, Yunnan 650500, People's Republic of China
\\
Key Laboratory of Survey Science of Yunnan Province, Yunnan University, Kunming, Yunnan 650500, People's Republic of China\\
Graduate School of Science, Division of Particle and Astrophysical Science, Nagoya University, Chikusa-Ku, Nagoya, 464-8602, Japan}
  
\email{shimabukuro@ynu.edu.cn}

\date{\today}

\begin{abstract}
We show that persistence-based topology of the 21\,cm forest encodes
information about Cosmic Dawn that is complementary to traditional
amplitude- or correlation-based statistics.
Applying topological data analysis to simulated one-dimensional forest
spectra over a grid of X-ray heating efficiencies $f_X$ and warm-dark-matter
masses $m_{\rm WDM}$ (which set the free-streaming scale), we construct
persistence diagrams and Betti--0 curves that track the birth--merger
hierarchy of absorption troughs under sublevel filtrations.
From these summaries we define three interpretable descriptors: the trough
line density $\lambda(t_\star)$, the total squared persistence
$M_2=\sum_{j\in \mathcal{I}_{\mathrm{long}}}\tau_j^2$
where $\mathcal{I}_{\mathrm{long}}\equiv\{\,j\mid\tau_j\ge\tau_\star\,\}$ denotes long-lived components),
and the Betti-curve asymmetry
$A_{\rm skew}$.
In a Fisher forecast around a fiducial WDM model, $\lambda(t_\star)$ and
$A_{\rm skew}$ provide strong local leverage on the heating axis, while
$M_2$ retains appreciable sensitivity to the free-streaming scale and
supplies an inclined constraint direction that reduces the remaining
degeneracy in the $(f_X,m_{\rm WDM})$ plane.
We further demonstrate that, under an SKA1-Low--like uncorrelated thermal-noise
model, noise predominantly produces short-lived fluctuations that are removed by imposing the same persistence-lifetime cut $\tau_j\ge\tau_\star$,
leaving the topology of long-lived troughs and
the gross Betti-curve morphology largely intact.
These results establish persistence-based descriptors as a robust non-Gaussian
probe of small-scale structure and heating during Cosmic Dawn, naturally
complementing power-spectrum and wavelet-based analyses.

\end{abstract}

\maketitle

\section{Introduction}

The 21\,cm forest consists of narrow absorption features imprinted on the
spectra of high-redshift radio sources by intervening neutral hydrogen
during Cosmic Dawn and the Epoch of Reionization
\citep{Carilli_2002,2002ApJ...579....1F,Furlanetto_2006,2013MNRAS.428.1755C,2015aska.confE...6C}.
These absorption structures trace the thermal and dynamical state of the
intergalactic medium, linking the observed brightness temperature to the
spin temperature, density fluctuations, and line-of-sight velocity
gradients. As next-generation facilities such as SKA1-Low approach the
sensitivity required for detecting the 21~cm forest \citep{2015aska.confE...6C},
identifying observables that can extract reliable physical information from its absorption morphology has become increasingly important.

A key difficulty is that astrophysical X-ray heating and dark-matter free-streaming can influence the forest in ways that look similar when viewed through traditional amplitude-based statistics. Heating raises the spin temperature toward the radiation temperature, flattening the contrast of absorption features, while dark matter with a finite free-streaming scale suppresses small-scale density fluctuations and reduces the number of absorption minima. When characterized using the one-dimensional power spectrum or related amplitude-weighted measures, these two mechanisms often lead to comparable reductions of small-scale power \citep{Xu_2009,Xu_2011,2014PhRvD..90h3003S,2016MNRAS.455..962S,2020PhRvD.101d3516S,2020PhRvD.102b3522S,Soltinsky_2021,2021JCAP...04..019K,2023PhRvD.107l3520S,2023PASJ...75S..33V,Shao_2024,2025arXiv251200402S}. Higher-order or multiscale statistics, including wavelet scattering transforms \citep{Thyagarajan_2020,Shao_2023,Soltinsky_2025,2024PhRvL.132w1002T,2025PhRvD.112f3557S}, extend sensitivity to nonlinear structure but remain fundamentally tied to fluctuation amplitudes. Recent AI/ML-based approaches have also demonstrated the ability to extract non-Gaussian information from 21\,cm observables beyond correlation-based statistics \citep{Sun_2024,Patil_2025}. These methods are promising, but their performance typically depends on the training domain and on how observational systematics and foreground or calibration residuals are represented in the training data. This motivates complementary summary statistics that are physically interpretable, can be stress-tested under controlled forward models, and target aspects of the absorption morphology not reducible to amplitudes alone.

From a physical standpoint, the two processes also differ in how they reshape absorption morphology: dark-matter free streaming eliminates shallow extrema and changes how neighboring troughs merge, whereas X-ray heating largely produces a monotonic remapping of the brightness-temperature field that tends to preserve the rank ordering of local minima. This suggests that summaries sensitive to the organization and merging hierarchy of troughs, rather than amplitudes alone, are natural candidates for separating heating from free streaming.

Topological data analysis (TDA) provides a principled framework for extracting this type of information \citep{2015arXiv150608903O,2016arXiv160908227W,2018arXiv180910745G}. In a one-dimensional field, sublevel filtrations track the emergence and merger of connected components as the threshold value rises from deep to shallow intensities. Persistence diagrams and Betti curves summarize these events by encoding the hierarchy of absorption troughs in terms of their birth values, lifetimes, and merger structure. These topological summaries capture the organization of minima and are invariant under smooth monotonic transformations of the field, a property that aligns closely with the qualitative effect of X-ray heating, while remaining sensitive to the removal of small-scale structure induced by dark-matter free streaming.

Topological approaches have already proven valuable in various cosmological contexts, including studies of the cosmic web, reionization morphology, and weak-lensing convergence fields \citep{2017MNRAS.465.4281P,2019MNRAS.486.1523E,2021MNRAS.507.2968W,2021A&A...648A..74H,2023MNRAS.520.2709E,2024MNRAS.529.4325B,2025JCAP...09..064C}. However, the absorption topology of the 21\,cm forest has not yet been explored. The forest is particularly well suited for such an analysis because its one-dimensional nature allows the full merger hierarchy of absorption features to be characterized without ambiguity, and because the narrow spectral structures expected during Cosmic Dawn are well resolved at kilohertz-level channel widths available to SKA1-Low.

In this work we introduce a topological framework for analyzing the 21\,cm forest and examine its potential to separate the physical effects of X-ray heating and dark-matter free-streaming. We intentionally focus on a minimal two-parameter testbed, $(f_X, m_{\rm WDM})$.
These parameters represent two qualitatively distinct ways of reshaping the one-dimensional
absorption morphology: an approximately monotonic remapping associated with heating,
and the removal or merging of shallow extrema associated with dark-matter free streaming.
This controlled setting provides a clean framework for assessing what persistence-based
topology captures beyond amplitude-based summaries, because persistence encodes the birth--merger hierarchy of troughs rather than amplitudes alone.
Other physical effects relevant to the 21~cm forest, such as Ly$\alpha$ coupling,
inhomogeneous reionization, redshift-space distortions, and detailed observational
systematics, are undoubtedly important but are deferred to future extensions of
the present proof-of-concept analysis. 
By focusing on the connectivity and rank structure of absorption troughs rather than on their amplitudes alone, topology offers a complementary perspective on the forest that is naturally robust to monotonic distortions and, in practice, can mitigate certain calibration and continuum residuals after appropriate preprocessing. This conceptual foundation provides a basis for incorporating topological information into future studies of the 21\,cm forest and for enhancing the physical insights that upcoming observations may deliver.


\section{Formalism of the 21~cm forest}

We analyze mock one-dimensional 21\,cm forest spectra generated from the semi-numerical simulations of \citet{Shao_2023}, which model the intergalactic medium (IGM) during Cosmic Dawn and the Epoch of Reionization for both cold and warm dark-matter cosmologies. These simulations incorporate radiative and X-ray heating processes with varying efficiencies and produce self-consistent density, spin-temperature, and ionization fields. In the following analysis, the thermal history of the IGM is controlled by the X-ray heating efficiency parameter $f_X$, which rescales the X-ray emissivity of early sources relative to a fiducial model, with larger values corresponding to more efficient heating. The nature of dark matter is parameterized by the warm-dark-matter particle mass $m_{\rm WDM}$, which sets the free-streaming scale and determines the degree of suppression of small-scale density fluctuations; the limit $m_{\rm WDM}\rightarrow\infty$ corresponds to the standard cold dark matter (CDM) scenario. 

In the 21~cm forest, the relevant observable is the attenuation of a bright background continuum source by intervening neutral hydrogen. In this regime, the observed flux density is
\begin{equation}
S_{\nu}^{\rm obs} = S_{\nu}\exp(-\tau_\nu),
\end{equation}
so that the optical depth $\tau_\nu$ directly governs the absorption features.

The corresponding 21\,cm optical depth along each line of sight is computed as
\begin{equation}
\begin{aligned}
\tau_{\nu}=\;&
\frac{3 h_{\rm P} c^3 A_{10} n_{\rm HI}}
     {16 k_{\rm B} T_S \nu_{21}^2 H(z)} \\
&\times
\left(1+\frac{1}{H(z)}\frac{dv_\parallel}{dr_\parallel}\right)^{-1},
\end{aligned}
\label{eq:tau}
\end{equation}
where $h_{\rm P}$ is Planck's constant, $c$ is the speed of light, $k_{\rm B}$ is Boltzmann's constant, and $A_{10}$ is the spontaneous Einstein coefficient of the 21\,cm transition. The quantity $n_{\rm HI}$ denotes the neutral hydrogen number density, $H(z)$ is the Hubble expansion rate at redshift $z$, and $dv_\parallel/dr_\parallel$ is the line-of-sight peculiar-velocity gradient along the comoving coordinate $r_\parallel$.

For later convenience, we express the same absorption in brightness-temperature units, which makes the dependence on the thermal state of the gas (through $T_S$) and the background radiation field explicit.
The observable 21\,cm forest signal can be written as the brightness-temperature imprint of H\,{\sc i} absorption on a background radiation field. In the optically thin limit, the differential brightness temperature is
\begin{equation}
\delta T_b(\nu) \approx \frac{T_S(z) - T_\gamma(z)}{1 + z} \, \tau_{\nu}(z),
\label{eq:Tb}
\end{equation}
where $\nu_{21}=1420.4\,{\rm MHz}$ is the rest-frame 21\,cm frequency, $T_S$ is the spin temperature of neutral hydrogen, and $T_\gamma$ is the background radiation temperature dominated by the continuum brightness temperature of the background radio source.
This relation provides a direct link between theoretical modeling of the IGM absorption (through $\tau_{\nu}$) and the observable brightness-temperature fluctuations.
The observing frequency $\nu$ is related to redshift by $\nu=\nu_{21}/(1+z)$.

Through Eqs.~\eqref{eq:tau} and \eqref{eq:Tb}, spatial variations in $T_S$, $n_{\rm HI}$, and the velocity field modulate $\tau_\nu$ and thus imprint narrow absorption features in $\delta T_b$, whose morphology carries information about local density structure and heating.

We explore a two-parameter space consisting of X-ray heating efficiencies
$f_X=\{0.0,0.1,0.2,0.3,0.5,1.0,2.0,3.0,5.0\}$ and warm-dark-matter masses
$m_{\rm WDM}=\{3,4,5,6,7,8,9,10\}\,{\rm keV}$,\linebreak
together with a cold dark matter (CDM) reference model.

For each model we analyze ten independent realizations, each providing ten
lines of sight spanning $10\,{\rm cMpc}$, sampled at $1\,{\rm kHz}$, matching
the expected channel width of SKA1-Low \citep{braun2019anticipatedperformancesquarekilometre}.
The choice of ten sightlines is motivated by theoretical estimates suggesting
that approximately twenty bright radio quasars may exist at $z \simeq 9$
\citep{2025ApJ...978..145N}, implying that our sample size is representative of
realistic observational prospects.

Before applying the topological analysis, each raw spectrum $x(\nu)$ is detrended and standardized. We adopt a Z-score normalization,
\begin{equation}
Z(\nu) = 
\frac{x(\nu) - \langle x\rangle}{\sigma_x},
\label{eq:Zscore}
\end{equation}
where $\langle x\rangle$ and $\sigma_x$ denote the mean and standard deviation along the same line of sight. This standardization removes differences in overall amplitude and rescales all models onto a common threshold axis, ensuring that the subsequent analysis is sensitive to morphological structure rather than absolute-amplitude variations.

\begin{figure*}[htbp]
\centering
\includegraphics[width=\linewidth]{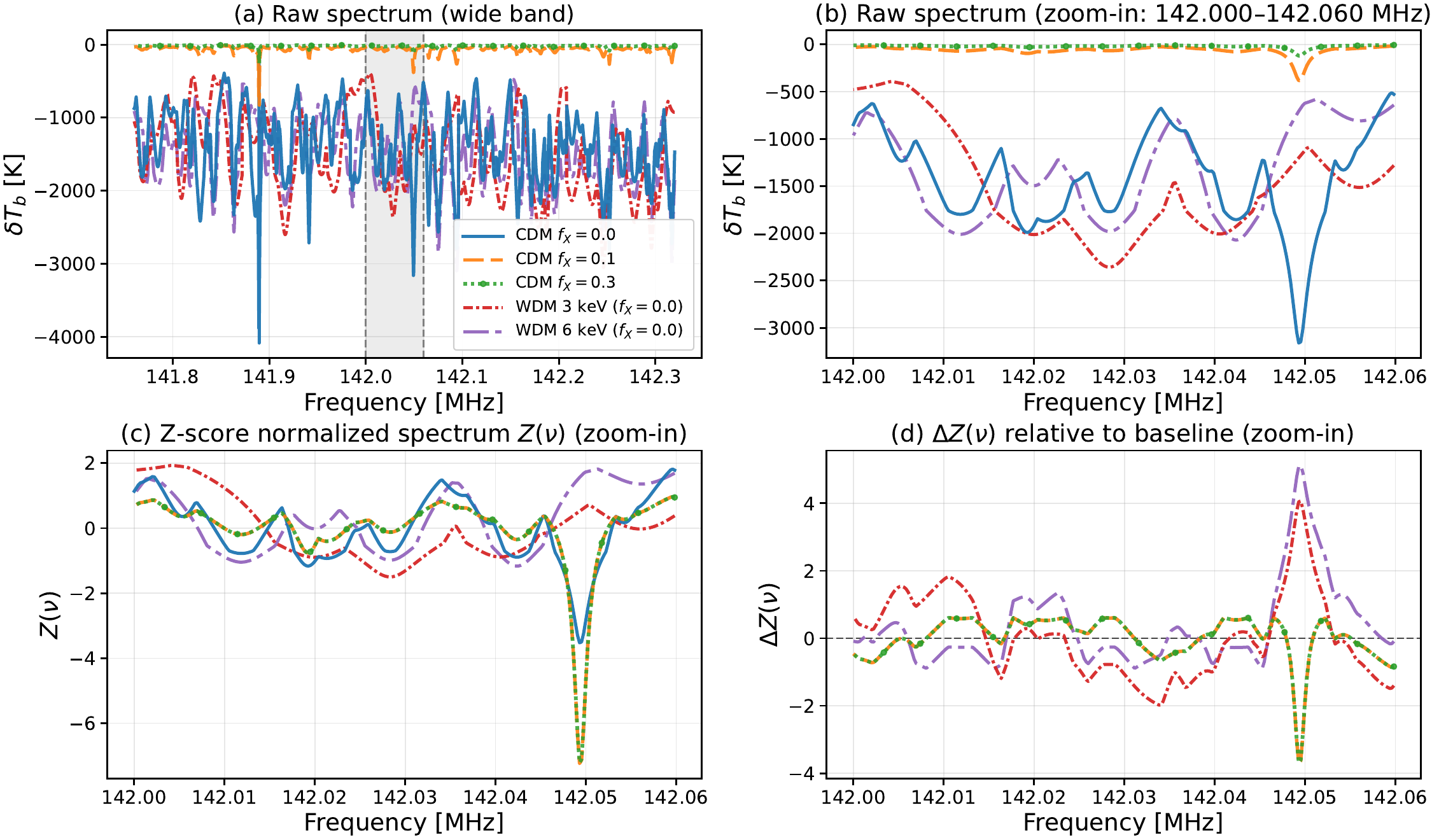}
\caption{Representative noise-free one-dimensional 21~cm forest spectra at $z\simeq 9$ from the same simulation realization and the same line of sight,
shown for CDM with varying X-ray heating efficiency ($f_X=0,\,0.1,\,0.3$) and WDM models with $m_{\rm WDM}=3,\,6$~keV (at $f_X=0$).
(a) Raw brightness-temperature spectra over the full frequency span.
(b) Zoom-in view over 142.000--142.060~MHz (Shaded region in (a)).
(c) The corresponding per-LOS standardized field $Z(\nu)$ defined in Eq.~\ref{eq:Zscore}
(d) $\Delta Z(\nu)\equiv Z_{\rm model}(\nu)-Z_{\rm ref}(\nu)$ relative to the baseline model (CDM, $f_X=0$); the dashed line indicates $\Delta Z=0$.}
\label{fig:forest_spectra_example}
\end{figure*}

Figure~\ref{fig:forest_spectra_example}  shows representative noise-free 21~cm forest spectra at $z\simeq 9$,
extracted from the same simulation realization and the same line of sight across the models considered in this work.
Panel (a) displays the raw brightness-temperature spectrum over the full frequency span, while panel (b) zooms into a narrow window
(142.000--142.060~MHz) to make individual absorption troughs more legible.
Because our topological analysis is applied to the detrended and per-LOS standardized field, panel (c) shows the corresponding
Z-score normalized spectrum $Z(\nu)$ defined in Eq.~(4), and panel (d) shows the difference
$\Delta Z(\nu)\equiv Z_{\rm model}(\nu)-Z_{\rm ref}(\nu)$ relative to the baseline model (CDM, $f_X=0$).
In this particular zoom window the two heated CDM cases ($f_X=0.1$ and $0.3$) nearly overlap in raw units, indicating that once heating is efficient the local spectral morphology can become close to saturated; the standardized and differential views in panels (c) and (d) therefore provide a common threshold axis on which to compare morphology across models.
These complementary views connect the qualitative discussion in Sec.~III to the spectra:
X-ray heating primarily reshapes the field in a smooth manner, whereas warm-dark-matter free streaming alters the small-scale trough pattern
through the suppression and merging of shallow extrema.
We emphasize that Fig.~\ref{fig:forest_spectra_example} is intended as a qualitative illustration.

\section{Topological formalism}

In a one-dimensional field such as $Z(\nu)$, topological information can be extracted by
tracking how absorption troughs appear, persist, and merge as an intensity threshold $t$ is swept across the brightness-temperature landscape.  Throughout this paper, we increase the threshold $t$ from more negative (deeper absorption) to less negative (shallower absorption) values. Concretely, for each $t$ we consider the connected regions where the field lies below the threshold, and follow how the connectivity of these regions evolves as $t$ increases. This threshold-sweep view provides a natural description of the birth--merger hierarchy of absorption features.

Because this construction is driven by the order in which extrema and saddle-like merge
events are encountered during the sweep, it depends primarily on the relative depth
hierarchy among pixels rather than on their absolute amplitudes. We refer to this property
as rank ordering: only the ordering of values from deeper to shallower matters, not
the exact numerical separations between them. A smooth monotonic remapping of the field, such as that induced by X-ray heating, preserves the rank ordering and therefore leaves the underlying connectivity evolution
essentially unchanged (up to a reparameterization of the threshold axis). In contrast,
dark-matter free streaming can erase or merge shallow extrema, thereby altering the set of
minima and their merger hierarchy.

Strictly speaking, the topology defined by this threshold-based construction is invariant
under smooth monotonic remappings of the field, provided thresholds are reparameterized
accordingly. In the language of topology, this is the invariance of the sublevel-set
filtration. In practice, however, our summary descriptors are evaluated with a common
persistence cut $\tau_\star$ and with a reference threshold defined for each line of sight
(e.g., the peak of the persistence-filtered Betti--0 curve). Under these conventions, even a
monotonic remapping can shift the Betti support along the threshold axis and thereby modify
shape-based descriptors, while changes that erase or merge extrema (such as free streaming)
alter the merger hierarchy itself.

\subsection{Sublevel filtrations and persistence}

For a one-dimensional field $Z(\nu)$, the sublevel set at threshold $t$ is
\begin{equation}
X_t = \{\,\nu\;|\; Z(\nu) \le t \,\}.
\end{equation}
In our analysis, $Z(\nu)$ denotes the detrended and per-LOS standardized 21~cm forest spectrum defined in Eq.~(4) (see Fig.~\ref{fig:forest_spectra_example}(c)).
As $t$ increases, new absorption troughs appear as
isolated connected components and later merge with deeper structures. This procedure defines a filtration $\{X_t\}_{t\in\mathbb{R}}$, and the evolution of connected components is encoded in
the zeroth homology group $\mathrm{H}_0(X_t)$. (Here $\mathrm{H}_0$ denotes homology and should not be confused with the Hubble constant $H_0$.)

\begin{figure}[htbp]
    \centering
    \includegraphics[width=1.0\hsize]{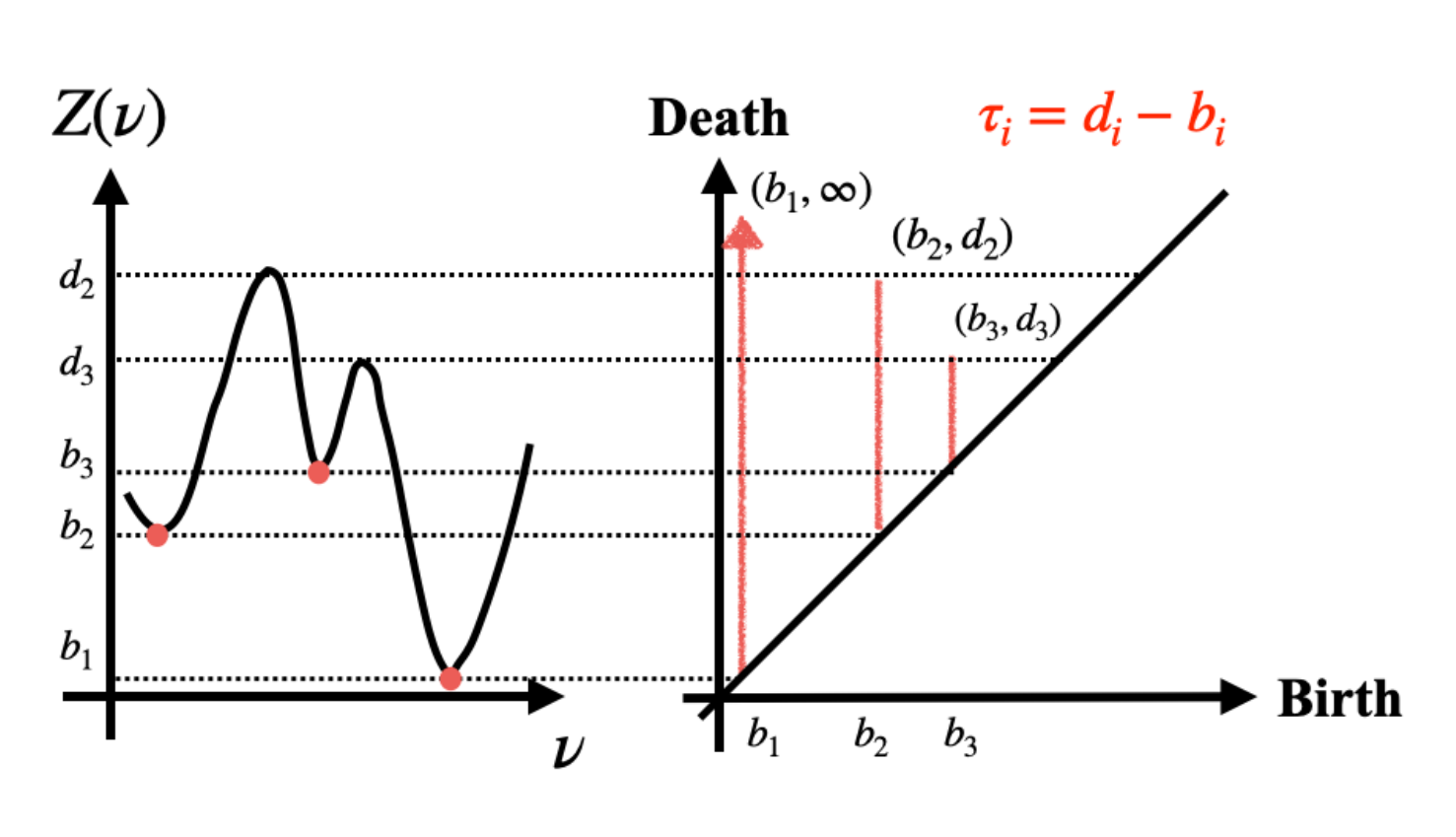}
    \caption{Schematic illustration of the sublevel-set filtration (left) and the corresponding persistence diagram (right) for a one-dimensional field $Z(\nu)$. Connected components are born at local minima ($b_i$) and merge at higher thresholds, yielding persistence pairs $(b_i,d_i)$ with lifetime $\tau_i=d_i-b_i$. The earliest-born component survives to all higher thresholds and therefore has no finite death. The pair $(b_1,\infty)$ is shown for illustration; in the data analysis, boundary-touching components are treated as boundary-censored and excluded from the persistence statistics.}

    \label{fig:pd_schematic}
\end{figure}

Each connected component has a birth threshold $b_j$ at which it appears,
and a death threshold $d_j$ at which it merges with an older (deeper) component.
The collection of points $(b_j, d_j)$ forms the persistence diagram (PD), and
the persistence lifetime
\begin{equation}
\tau_j = d_j - b_j
\end{equation}
quantifies the prominence of an absorption trough. Hereafter, $\tau$ denotes the persistence lifetime, not the optical depth. Long-lived features correspond to deep, well-separated troughs, while short-lived features track shallow fluctuations that are easily erased by noise or small-scale smoothing.  Figure~\ref{fig:pd_schematic} provides a schematic illustration of this construction for a
one-dimensional field, showing how absorption troughs are born and merge under a threshold
sweep and how these events are mapped onto the persistence diagram.

A complementary summary is the Betti--0 curve,
\begin{equation}
\beta_0(t)=\#\{\,j\; |\; b_j \le t < d_j\,\},
\end{equation}
which counts the number of connected absorption components alive at threshold $t$.
Betti--0 therefore traces how many distinct troughs remain separated at a given depth
and provides a direct measure of the merger hierarchy of potential wells in the IGM.

\begin{figure}[htbp]
  \centering
  \includegraphics[width=1.0\linewidth]{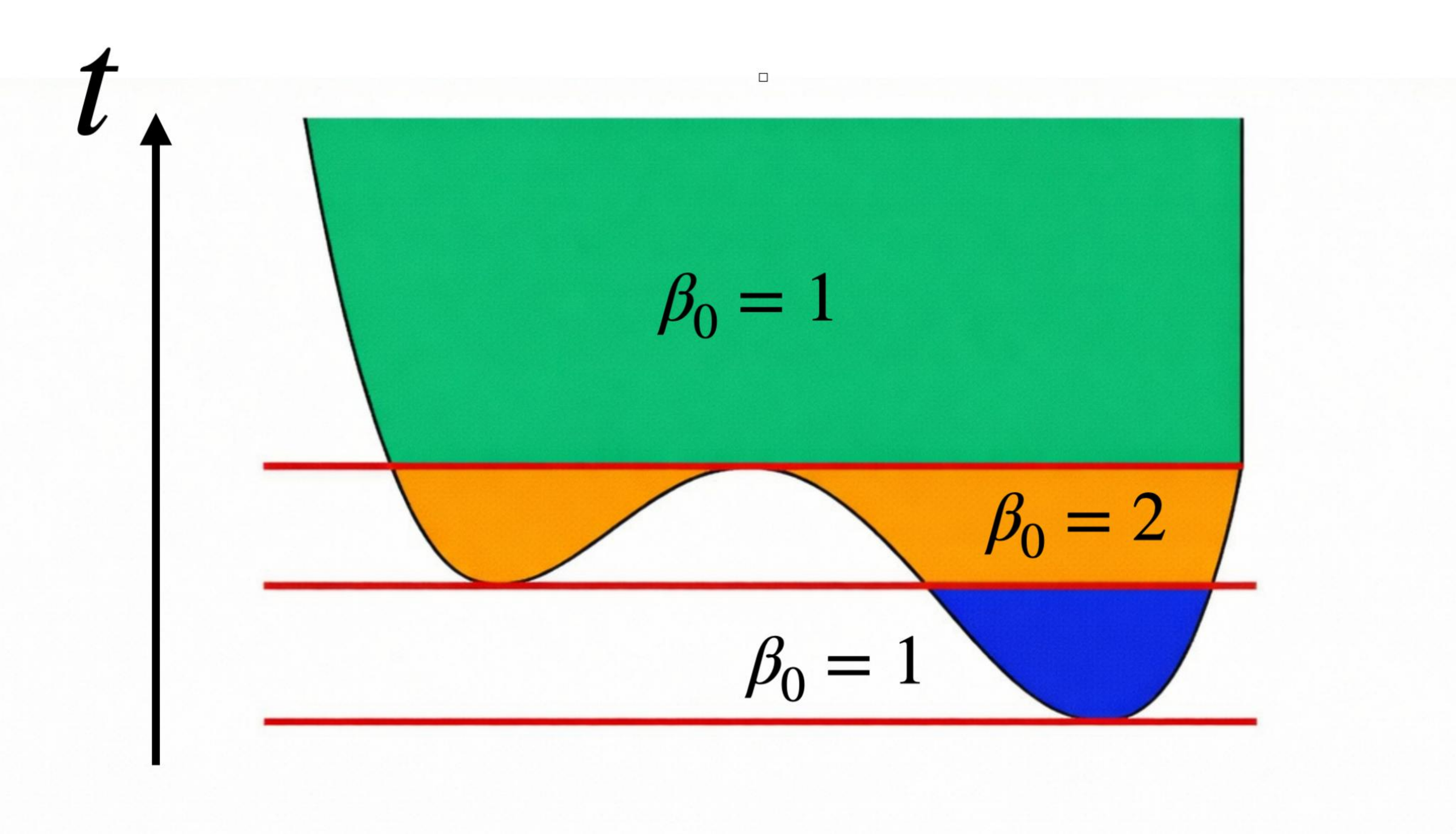}
  \caption{
  Schematic illustration of a sublevel-set filtration for a one-dimensional absorption field.
  As the threshold $t$ is increased, disconnected absorption troughs
  appear and subsequently merge. The Betti--0 number $\beta_0(t)$ counts the number of connected components at each threshold and changes discretely when merger events occur. This figure is purely illustrative and is intended to clarify the definition of $\beta_0(t)$.This schematic applies to the standardized 1D field $Z(\nu)$ used in our filtration.
}
  \label{fig:betti_schematic}
\end{figure}

Having defined the Betti--0 curve, it is useful to clarify its relation to Minkowski functionals,
which are widely used to quantify morphology in cosmology. Both approaches start from thresholded
excursion sets: given a threshold $t$, one studies the geometry and connectivity of the set
$X_t=\{\nu\,|\,Z(\nu)\le t\}$ (or, equivalently, superlevel sets).
In one dimension, the Minkowski description reduces essentially to the measure (total length)
of $X_t$ and its Euler characteristic $\chi(X_t)$.
Because one-dimensional excursion sets have no loops, $\chi(X_t)$ is equal to the number of
connected components, and therefore coincides with the Betti--0 number, $\chi(X_t)=\beta_0(t)$.
In this sense, the Betti--0 curve can be viewed as the connectivity part of the Minkowski
description evaluated as a function of threshold.
The persistence diagram extends this picture by recording when each connected component is
created and when it merges as the threshold varies, thereby encoding the full merger hierarchy
of troughs across thresholds rather than a snapshot at each $t$.
In the following, we compute persistence diagrams and Betti--0 curves from the standardized spectra $Z(\nu)$ and use them to construct the three descriptors introduced in Sec.~III.C, providing a direct link between the schematic constructions in Figs.~\ref{fig:pd_schematic} and \ref{fig:betti_schematic} and the model-dependent spectra shown in Fig.~\ref{fig:forest_spectra_example}. 

Figure~\ref{fig:betti_schematic} illustrates this construction schematically for a
one-dimensional absorption field.
As $t$ increases, new connected components
are created at local minima of $Z(\nu)$, while existing components merge when the threshold
reaches higher saddle points.
The Betti--0 number $\beta_0(t)$ therefore counts the number of connected components in the
sublevel set $X_t$ at each threshold, encoding the birth--merger hierarchy of absorption
troughs along the line of sight.

Figures~\ref{fig:pd_schematic} and \ref{fig:betti_schematic} are schematic illustrations. In the actual analysis of simulated 21~cm forest spectra, the signal is defined on a finite frequency window.
We therefore adopt an open-boundary (boundary-censored) convention: connected components that intersect either edge of the window are treated as neither births nor deaths.
Operationally, such edge-touching features are excluded from the persistence statistics, and we compute $\beta_0(t)$ and all derived descriptors using only components whose birth and merger events occur within the interior of the window.

\subsection{Persistence filtering and reference thresholds}

The raw PD contains many short-lived features associated with shallow or noise-like fluctuations.
To isolate physically meaningful troughs, we impose a persistence cut
\begin{equation}
\tau_\star = 0.411,
\end{equation}
defined as the 95th percentile of the persistence-lifetime distribution in the fiducial
(noiseless CDM) model, computed from all components across all LOS.
This defines the set of long-lived components
\begin{equation}
\mathcal{I}_{\mathrm{long}} = \{\, j \mid \tau_j \ge \tau_\star \,\}.
\end{equation}
The precise value of $\tau_\star$ is not essential; varying it within reasonable bounds
preserves all qualitative trends. Unless stated otherwise, whenever a descriptor requires an
intensity threshold we define a line-of-sight--dependent reference value
$t_{\star,\ell}\equiv\arg\max_t\,\beta_{0,\ell}(t)$ from the persistence-filtered Betti--0 curve of
that LOS, and we report LOS-averaged descriptor values. For notational simplicity, we often omit
the LOS index and write $t_\star$. This peak-based convention avoids imposing an external absolute
threshold but implies that model-dependent shifts of the Betti support along the threshold axis
can contribute to the response of $\lambda$ and $A_{\rm skew}$ in addition to changes in the
underlying merger hierarchy.

\subsection{Topological descriptors}

From the PD we construct three physically interpretable descriptors that capture
complementary aspects of the absorption morphology. Together, these quantities summarize how many independent troughs exist, how broadly their
lifetimes are distributed, and whether the lifetime distribution is dominated by long-
or short-lived components.
These descriptors respond in distinct ways to heating and dark-matter free-streaming,
thereby helping to break their degeneracy. Rather than working directly with the full persistence diagram, we compress its information into a small set of summary descriptors that are robust, physically interpretable, and
amenable to statistical inference.

\begin{enumerate}[label=(\roman*)]
\item {\it Trough line density}.
For a frequency span $\Delta\nu$ at redshift $z$, the corresponding comoving line-of-sight
length is
\begin{equation}
L_\parallel=\Delta\nu\,\frac{d\chi}{d\nu},\qquad
\frac{d\chi}{d\nu}=\frac{c(1+z)^2}{H(z)\,\nu_{21}}.
\end{equation}
The line density of absorption troughs is then
\begin{equation}
\lambda(t_\star) = \frac{\beta_0(t_\star)}{L_\parallel}\;\;[{\rm Mpc^{-1}}],
\end{equation}
which measures the abundance of distinct troughs at the fiducial threshold.
Warm dark matter suppresses small-scale structure and therefore reduces $\lambda$,
while X-ray heating also decreases $\lambda$ by homogenizing the temperature field.

\item {\it Total squared persistence.}
To capture the "strength" of the absorption features, we compute the sum of the squared lifetimes
for all long-lived components ($\tau_j \ge \tau_\star$). We define the second moment sum as
\begin{equation}
M_{2} = \sum_{j \in \mathcal{I}_{\mathrm{long}}} \tau_{j}^{2}.
\label{eq:M2}
\end{equation}
Unlike a simple count, $M_{2}$ weights features by their prominence (lifetime).
Since X-ray heating compresses the contrast of absorption troughs (reducing $\tau_j$),
$M_{2}$ decreases rapidly with increasing $f_X$, making it a sensitive probe of the thermal state
of the IGM. At the same time, warm dark matter suppresses small-scale extrema and shortens the overall
lifetime distribution, leading to a systematic reduction of $M_2$ even in the absence of
strong heating.

\item {\it Betti curve asymmetry.}
Finally, we characterize the shape of the brightness-temperature distribution by quantifying the
asymmetry of the Betti-0 curve $\beta_0(t)$. The support of $\beta_0(t)$ reflects the depth range
of absorption features.
Let $t_{p}$ be the threshold value corresponding to the $p$-th percentile of the cumulative area
under the $\beta_0(t)$ curve, defined such that
$\int_{-\infty}^{t_p} \beta_0(t') dt' = \frac{p}{100} \int_{-\infty}^{\infty} \beta_0(t') dt'$.
Using the peak position $t_{\star}$ as a reference, we define the asymmetry factor
$A_{\mathrm{skew}}$ as the ratio of the widths of the shallow (right) and deep (left) sides of
the distribution:
\begin{equation}
A_{\mathrm{skew}} = \frac{t_{95} - t_{\star}}{t_{\star} - t_{05}}.
\label{eq:Askew}
\end{equation}
Here, $t_{95}$ characterizes the boundary of shallow fluctuations, while $t_{05}$ marks the deep
absorption tail. The use of percentile-based widths ensures that $A_{\mathrm{skew}}$ is insensitive to the
absolute normalization of $\beta_0(t)$ and robust against isolated outliers in the
distribution.

X-ray heating acts, to leading order, as a monotonic remapping of the field intensity, which can
distort the shape of $\beta_0(t)$ and alter the balance between the deep and shallow wings.
In contrast, warm dark matter primarily suppresses the abundance of features while preserving the
relative shape of the surviving distribution more effectively. This difference renders
$A_{\mathrm{skew}}$ primarily sensitive to $f_X$ (orthogonal to $\lambda$).

\end{enumerate}

\section{Results}

\subsection{Persistence lifetime distributions}

\begin{figure}[htbp]
  \centering
  \includegraphics[width=1.0\hsize]{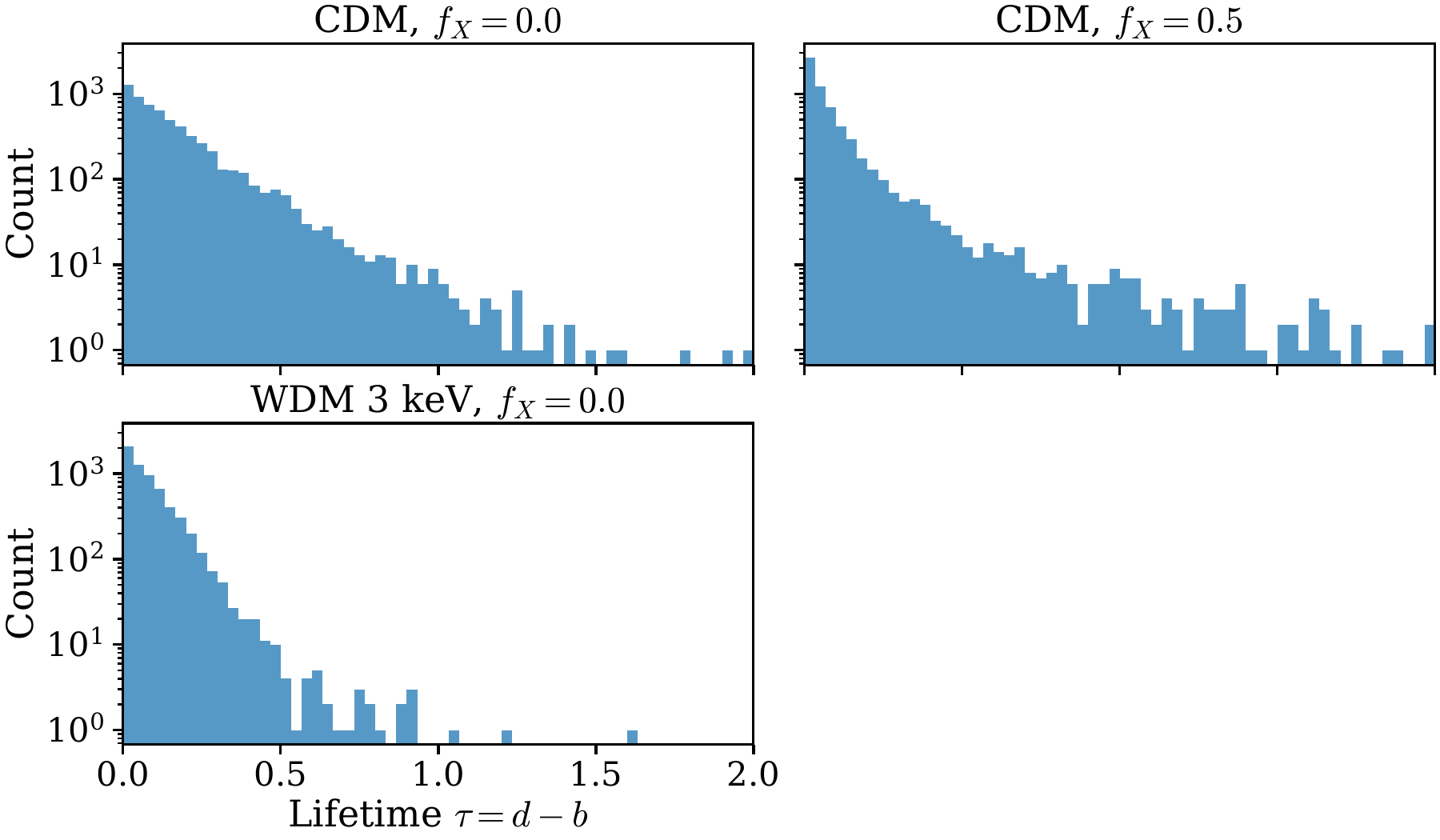}
  \caption{Distributions of persistence lifetimes(Death-Birth).
  Panels compare CDM with $f_X=0$ and $0.5$  and CDM ($f_X=0$) versus WDM ($m_{\rm WDM}=3$\,keV, $f_X=0$).}
  \label{fig:lifetime}
\end{figure}

Figure~\ref{fig:lifetime} compares the distributions of persistence lifetimes $\tau=d-b$ using common bins across models.
For CDM, increasing $f_X$ concentrates the distribution toward very small lifetimes ($\tau\simeq 0$) and suppresses the intermediate-lifetime population, consistent with X-ray heating smoothing weak, small-scale trough structure so that many components merge shortly after birth in the filtration. For WDM ($m_{\rm WDM}=3\,\mathrm{keV}$, $f_X=0$), the distribution is more strongly shifted toward short lifetimes and the population of long-lived features is reduced, reflecting the suppression of small-scale extrema and earlier merging of neighboring troughs.

Because the extreme long-$\tau$ tail is sparsely populated in these histograms, we characterize tail behavior using an integrated lifetime-weighted summary, $M_2 = \sum_{j\in I_{\rm long}} \tau_j^2$, rather than a raw count above the cut.

\subsection{Betti--0 curves}

\begin{figure}[htbp]
  \centering
  \includegraphics[width=1.0\hsize]{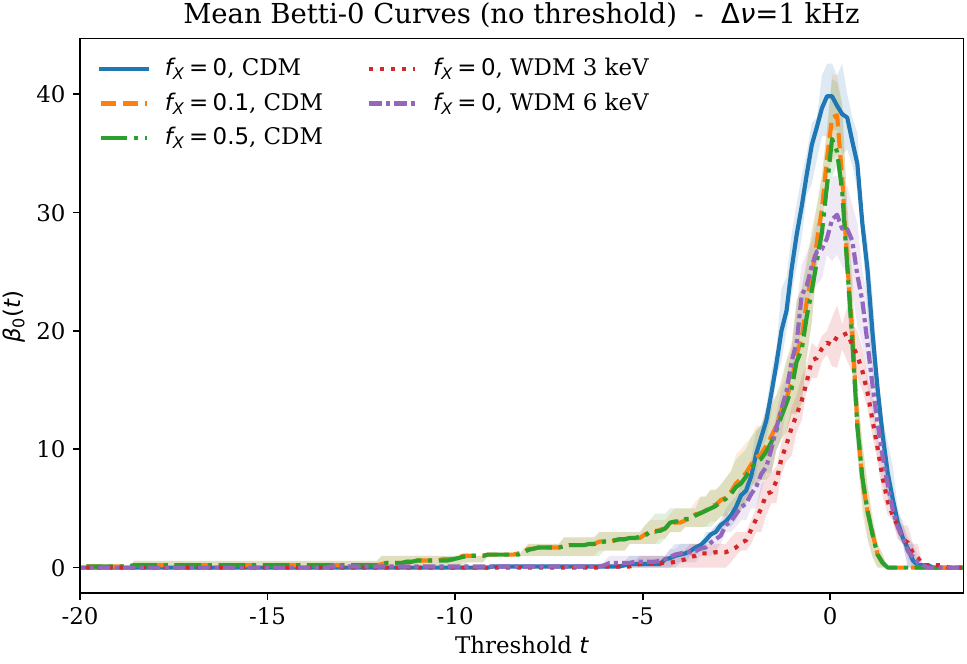}
  \includegraphics[width=1.0\hsize]{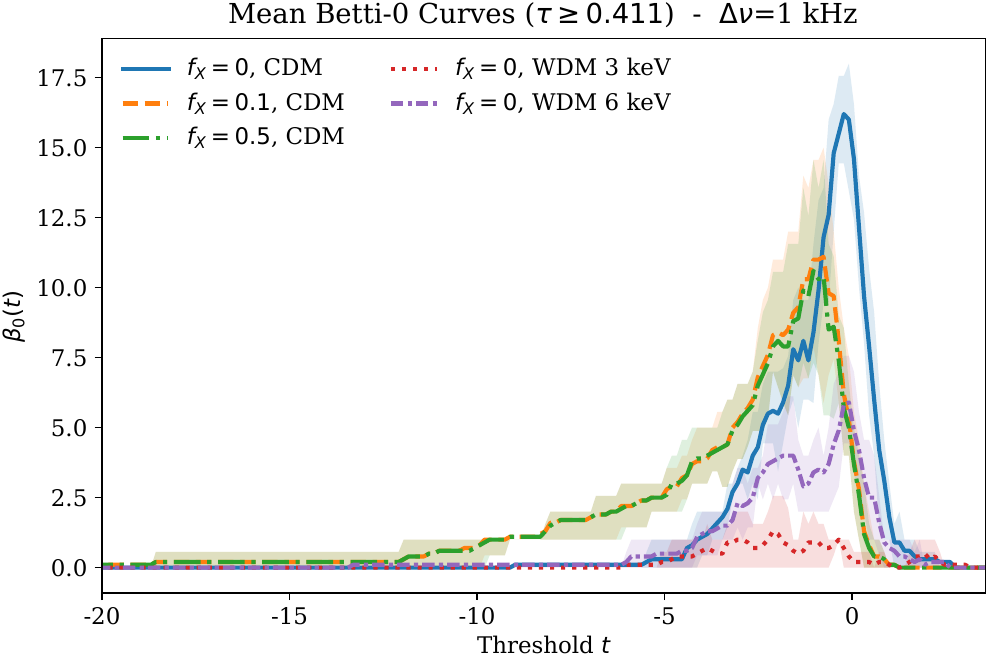}
  \caption{Mean Betti--0 curves $\langle\beta_0(t)\rangle$ at $\Delta\nu=1$\,kHz.
Top: without persistence filtering.
Bottom: after applying a uniform persistence cut $\tau\ge0.411$, which suppresses short-lived features and clarifies model differences.
Shaded regions show the central 68\% interval across lines of sight.}
  \label{fig:betti}
\end{figure}

Figure~\ref{fig:betti} presents the Betti--0 curves, i.e., the number of connected components
$\beta_0(t)$ alive at threshold $t$, for standardized brightness--temperature spectra at a spectral
resolution of 1\,kHz. We focus first on the persistence-filtered curves (bottom panel), which
isolate the topologically significant absorption troughs.

With persistence filtering applied, $\beta_0(t)$ exhibits a single pronounced peak as a function
of $t$. The rise toward the peak reflects the appearance of distinct troughs as the threshold
enters the typical absorption depths, whereas the decline beyond the peak reflects the progressive
merging of neighboring troughs as the threshold is increased. The peak height therefore encodes
the abundance of independent absorption features, while the peak width characterizes the range of
thresholds over which these features remain distinct.

For the unheated CDM case ($f_X=0$), $\beta_0(t)$ shows a high and broad peak, indicating many
absorption troughs that remain separate over a wide threshold range. Increasing the X-ray heating
efficiency reduces and narrows the peak, consistent with contrast compression and a more homogeneous
absorption landscape. The WDM model ($m_{\rm WDM}=3~{\rm keV}$, $f_X=0$) yields a smaller peak, consistent with the suppression of small-scale extrema and earlier merging of surviving troughs. Notably, this relative ordering differs from what is often emphasized by amplitude-based summaries. Many amplitude-based statistics respond most strongly to X-ray heating because heating raises the spin temperature and compresses the overall absorption contrast,
thereby reducing the fluctuation amplitude in a largely global manner. In contrast, warm-dark-matter
free streaming primarily suppresses small-scale density extrema and changes the merger hierarchy of
neighboring troughs, which can have a comparatively smaller imprint on purely amplitude-based
measures while leaving a clear signature in connectivity-based statistics. The Betti--0 response
therefore highlights the impact of free streaming on the connectivity evolution, illustrating the
complementary information captured by topology.

To clarify the role of persistence filtering, the top panel shows the unfiltered Betti--0 curves.
Without filtering, the raw curves include numerous short-lived components associated with weak
fluctuations and noise-like features, which introduce small oscillations and artificially broaden
the $\beta_0(t)$ profiles, partially obscuring the underlying cosmological trends. We therefore
apply a uniform persistence threshold $\tau \ge 0.411$, chosen as the 95th percentile of the
lifetime distribution in the baseline noiseless CDM model, to remove low-persistence features while
retaining long-lived absorption troughs that reflect the IGM morphology.

Overall, the comparison between unfiltered and filtered analyses demonstrates that persistence
filtering isolates the robust topological backbone of the 21\,cm forest. The adopted 1\,kHz
resolution ensures that the narrow absorption features expected in SKA1-Low observations are
sufficiently resolved to capture their intrinsic topological structure.

\subsection{Impact of observational effects} 

\begin{figure*}[htbp]
  \centering
  \includegraphics[width=\hsize]{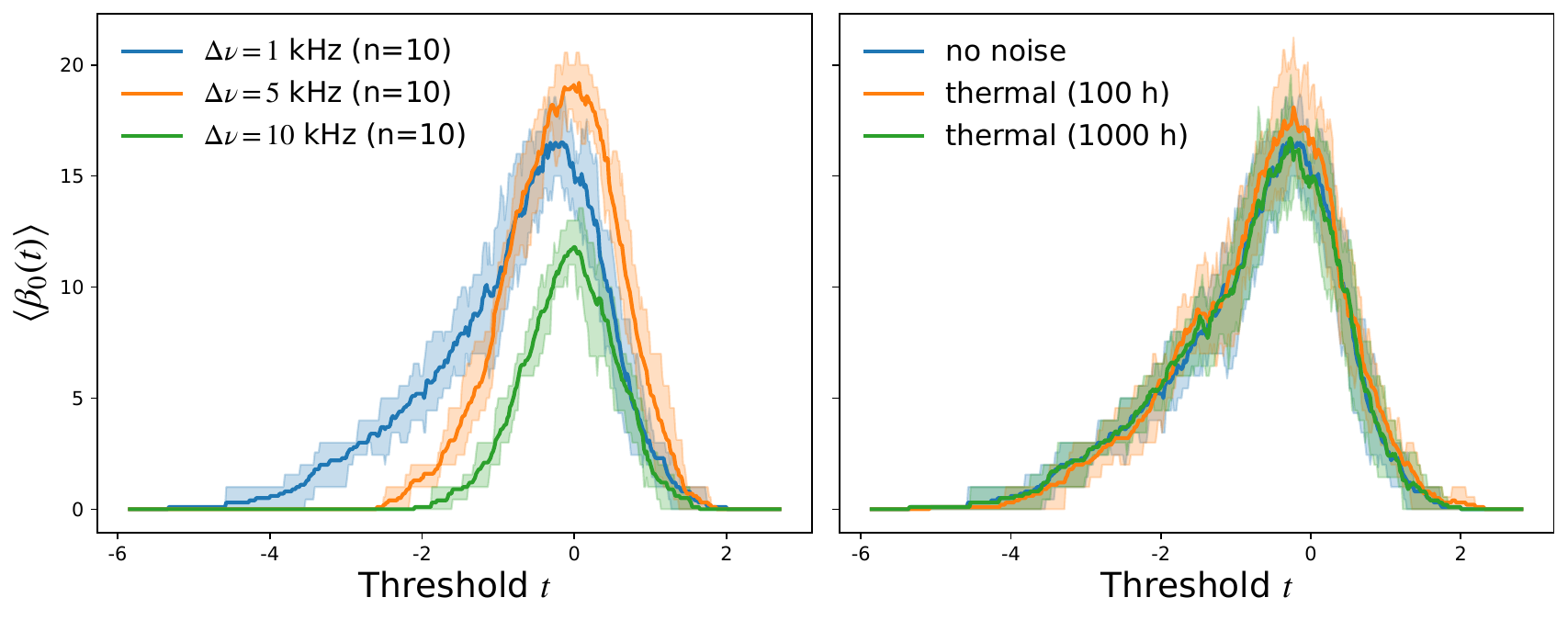}
  \caption{
{\bf Topological sensitivity for CDM ($f_X=0$).} A common persistence cut $\tau\ge0.411$ is used.
{\bf Left:} Resolution dependence (1, 5, 10\,kHz) shown without thermal noise.
{\bf Right:} Thermal noise at 1\,kHz (noiseless, $t_{\rm int}=100$ and $1000$\,h): increasing
$t_{\rm int}$ lowers the noise level, reduces LOS scatter, and brings the mean
$\beta_0(t)$ closer to the noiseless case.
}

  \label{fig:betti_res_noise}
\end{figure*}

We investigate how spectral resolution and thermal noise affect the topology of the 21\,cm forest
in the fiducial CDM model with $f_X=0$, as summarized in Figure~\ref{fig:betti_res_noise}.
We apply the same analysis pipeline to all spectra, consisting of detrending, per-LOS
standardization, downsampling, sublevel filtration, and Betti--0 evaluation, using a fixed
persistence threshold $\tau \ge 0.411$.
The left panel shows the dependence on spectral resolution in the absence of thermal noise by
varying the channel width $\Delta\nu$.
Because the descriptors are evaluated after applying a common lifetime cut, the persistence-filtered
Betti--0 response need not vary monotonically with $\Delta\nu$: spectral averaging can suppress
short-lived fluctuations while simultaneously increasing the typical lifetimes of the surviving
troughs, thereby changing how many components pass the fixed persistence threshold.

To model instrumental thermal noise, we adopt the radiometer equation and add Gaussian noise independent in each frequency channel. The rms noise in flux density is

\begin{equation}
\sigma_S(\nu)=\frac{\mathrm{SEFD}(\nu)}{\sqrt{n_{\rm pol}\,t_{\rm int}\,\Delta\nu}},
\end{equation}
where $\mathrm{SEFD}(\nu)=2k_{\rm B}T_{\rm sys}(\nu)/A_{\rm eff}(\nu)$, $n_{\rm pol}=2$ is the number
of polarizations, $t_{\rm int}$ is the integration time per background source, and $\Delta\nu$ is the
channel width. For reproducibility, in this work we adopt a fiducial constant sensitivity ratio
$A_{\rm eff}/T_{\rm sys}=800~{\rm m^2\,K^{-1}}$, which is equivalent to
$\mathrm{SEFD}=2k_{\rm B}/(A_{\rm eff}/T_{\rm sys})\simeq 3.45~{\rm Jy}$.
At $\Delta\nu=1~{\rm kHz}$ this yields representative per-channel noise levels
$\sigma_S\simeq 0.129~{\rm mJy}$ for $t_{\rm int}=100~{\rm h}$ and
$\sigma_S\simeq 0.0407~{\rm mJy}$ for $t_{\rm int}=1000~{\rm h}$. To model thermal-noise contamination under SKA1-Low--like observing conditions, we adopt the
instrumental sensitivity and system parameters described in Ref.~\cite{braun2019anticipatedperformancesquarekilometre}.

In a fully self-consistent observing model, the thermal-noise level decreases with increasing channel width as $\sigma_S\propto(\Delta\nu)^{-1/2}$. In this work we intentionally decouple this dependence by holding the noise model fixed while varying $\Delta\nu$, in order to isolate the morphological effect of spectral averaging.

We convert the flux-density noise $\sigma_S(\nu)$ into the corresponding relative noise level in
the absorbed spectrum by normalizing to the background-source continuum,
$\sigma_{\rm rel}(\nu)=\sigma_S(\nu)/S_\nu$.
We adopt a fiducial power-law continuum for the background source,
\begin{equation}
S_\nu = S_{150}\left(\frac{\nu}{150~\mathrm{MHz}}\right)^{\eta},
\end{equation}
with $S_{150}=10~\mathrm{mJy}$ and $\eta=-1.05$.
This fiducial spectrum is used as a representative (hypothetical) bright high-redshift radio source;
results for other sources can be rescaled as $\sigma_{\rm rel}\propto 1/S_\nu$.
With the above choices, the relative noise level at 150~MHz is
$\sigma_{\rm rel}\simeq 1.29\times10^{-2}$ (100~h) and $4.07\times10^{-3}$ (1000~h).
A Gaussian random realization with variance $\sigma_{\rm rel}^2(\nu)$ is added prior to applying the
same detrending and per-LOS standardization used throughout. Finally, since the absorbed spectrum is
often modeled as $F=\exp(-\tau_\nu)$, for $\tau_\nu\ll1$ we have $\delta\tau_\nu \simeq \delta F/F \simeq \sigma_{\rm rel}$,
so the quoted noise levels correspond to representative optical-depth uncertainties of order
$\delta\tau_\nu\sim10^{-2}$ (100~h) and a few $\times10^{-3}$ (1000~h).

The left panel of Figure~\ref{fig:betti_res_noise} shows the resulting resolution-only trend,
obtained by varying $\Delta\nu$ at fixed noise level to isolate the morphological effect of spectral averaging. The right panel of Figure~\ref{fig:betti_res_noise} then illustrates the impact of thermal noise at fixed $\Delta\nu=1$\,kHz, showing a
noiseless reference case and thermal-noise realizations corresponding to $t_{\rm int}=100$ and
$1000$\,h per background source.
Thermal noise introduces numerous short-lived, shallow fluctuations that can artificially increase
$\beta_0(t)$ at high thresholds, but applying the same persistence cut efficiently suppresses these
spurious components and isolates long-lived absorption structures.
As $t_{\rm int}$ increases, the Betti--0 curve converges toward the noiseless case and the
line-of-sight scatter decreases, demonstrating that the intrinsic topological signal remains
detectable under realistic SKA-like noise levels.

We focus on spectral resolution and \emph{uncorrelated} thermal noise because they are the most immediate, instrument-driven effects that act directly on one-dimensional forest spectra at the channel level and can be cleanly parameterized by $\Delta\nu$ and $t_{\rm int}$.
Other effects that can matter in practice include bandpass/calibration residuals (often with frequency correlations), continuum-fitting/detrending uncertainties, and irregular sampling or gaps due to RFI excision.
A fully end-to-end forward model should incorporate these effects and marginalize over associated nuisance parameters; see Sec.~V.F for further discussion.

\subsection{Topological degeneracy maps}

\begin{figure*}[t]
  \centering
  \includegraphics[width=\hsize]{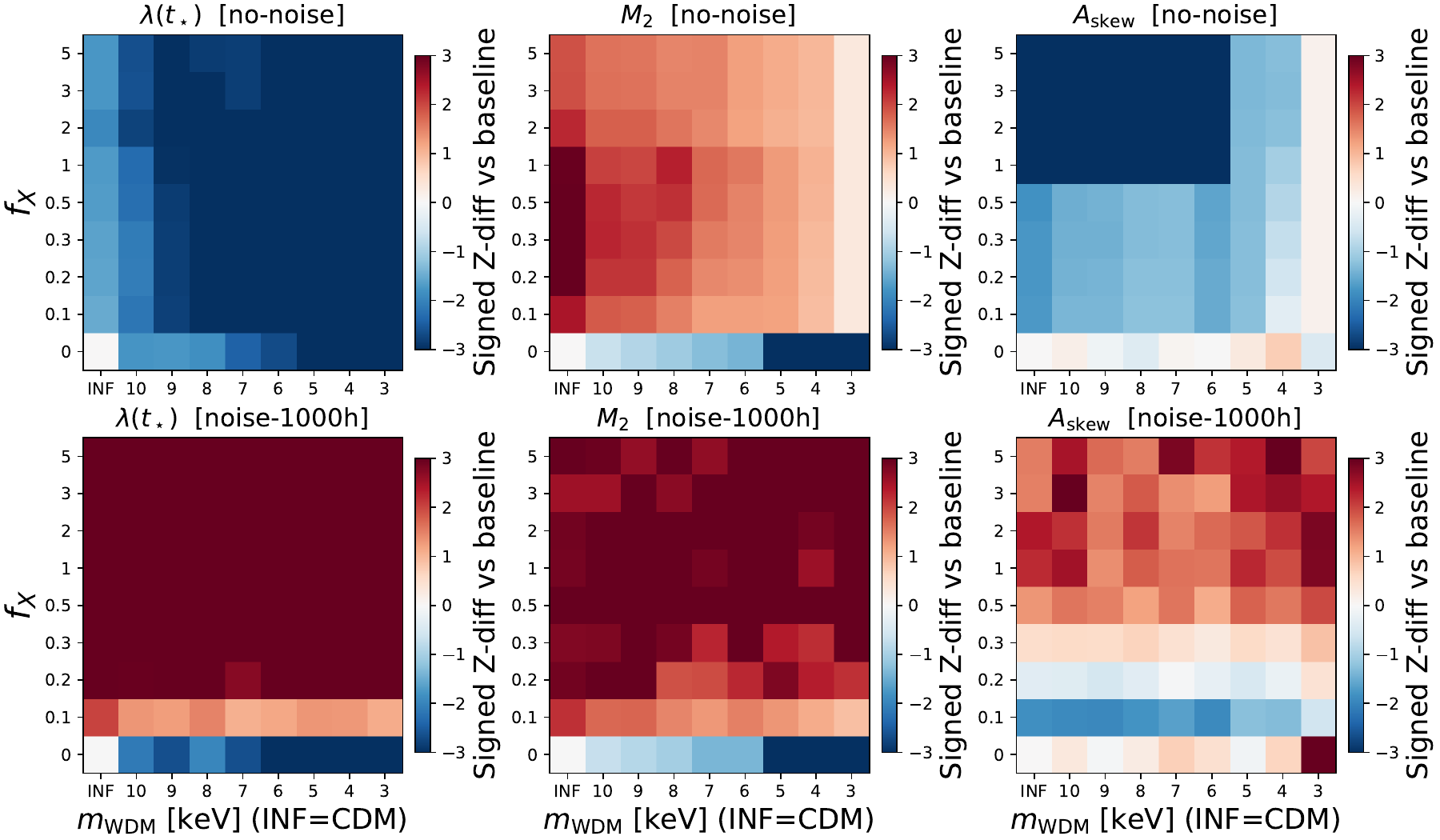}
  \caption{Topological descriptor maps across $(f_X,m_{\rm WDM})$, shown as signed differences relative to the baseline CDM model ($f_X=0$). Columns show $\lambda(t_{\star,\ell})$, $M_2$, and $A_{\rm skew}$. The top row is noiseless and the bottom row includes SKA1-Low thermal noise with $t_{\rm int}=1000$\,h. Persistence-based quantities are computed using a common lifetime cut $\tau\ge\tau_\star$.}

  \label{fig:tda_feature_heatmaps}
\end{figure*}

Figure~\ref{fig:tda_feature_heatmaps} summarizes the global trends of the three descriptors
using signed Z--difference maps relative to the baseline (CDM, $f_X=0$).  For each model we
compute the LOS-averaged descriptor value $\mu$ and an associated LOS scatter estimated from the 16--84 percentile range, and form a signed effect-size relative to the baseline.  For visualization, for each model we compute the LOS-averaged descriptor value $\mu_\alpha$ and its LOS scatter
$\sigma_{\rm LoS}$, estimated from the 16--84 percentile range as $\sigma_{\rm LoS} \equiv (P_{84}-P_{16})/2$.
We then define the signed standardized difference relative to the baseline model as
\begin{equation*}
Z_{\mathrm{diff},\alpha}(\boldsymbol{\theta}) \equiv
\frac{\mu_\alpha(\boldsymbol{\theta})-\mu_{\alpha,\mathrm{ref}}}{\sigma_{\alpha,\mathrm{ref}}}\, ,
\end{equation*}
where $\mu_{\alpha,\mathrm{ref}}$ and $\sigma_{\alpha,\mathrm{ref}}$ are the baseline mean and LOS scatter.
This $Z_{\mathrm{diff}}$ is an effect-size measure used only for compact visualization (not a statistical significance),
and it should not be confused with the standardized field $Z(\nu)$ defined in Eq.~\ref{eq:Zscore}.

All descriptors use the same persistence cut $\tau_\star$, while quantities involving the
threshold are evaluated at a line-of-sight--dependent reference level
$t_{\star,\ell}\equiv\arg\max_t\,\beta_{0,\ell}(t)$ obtained from the persistence-filtered
Betti--0 curve.  Consequently, the sign of the Z--difference reflects not only changes in the
abundance and lifetimes of robust troughs, but also model-dependent shifts in the placement and
shape of $\beta_0(t)$ that move the peak position across LOS.

The trough line density $\lambda(t_{\star,\ell})=\beta_{0,\ell}(t_{\star,\ell})/L_\parallel$
measures the number density of connected components at the characteristic threshold where each
LOS attains maximal connectivity.  In the noiseless case, decreasing $m_{\rm WDM}$ suppresses
small-scale extrema and typically lowers the peak height of $\beta_0(t)$, leading to smaller
$\lambda$ than in the baseline model.  With thermal noise, many short-lived fluctuations are
removed by the persistence cut, and the remaining Betti support and its peak position can shift
across LOS; this can weaken the apparent dependence of $\lambda$ on $m_{\rm WDM}$ and enhance
the relative role of $f_X$ through changes in the Betti-curve morphology.

The total squared persistence
$M_2=\sum_{j\in\mathcal{I}_{\rm long}}\tau_j^2$
captures the cumulative prominence of long-lived troughs selected by the uniform lifetime cut.
Increasing $f_X$ tends to shorten lifetimes by compressing contrast, while free streaming
reduces the abundance of small-scale absorbers and reshapes the lifetime distribution.
Accordingly, $M_2$ responds to both parameters and provides a sensitivity direction
complementary to those of $\lambda$ and $A_{\rm skew}$.

The Betti-curve asymmetry $A_{\rm skew}$ quantifies whether the support of $\beta_0(t)$ is
weighted toward deep or shallow thresholds.  Because it depends on the relative placement of the
Betti support along the threshold axis, $A_{\rm skew}$ is particularly sensitive to heating-driven
remappings that distort the balance between the deep and shallow wings, providing information
distinct from the primarily count-based $\lambda$ and the lifetime-weighted $M_2$.

Thermal noise generates many low-persistence fluctuations, but these are efficiently suppressed by
the uniform lifetime threshold.  The residual large-scale gradients in
Fig.~\ref{fig:tda_feature_heatmaps} therefore, trace robust, long-lived absorption structures and the
shape of their Betti support, providing a compact visualization of how heating and free streaming
imprint complementary signatures in the descriptor space.

\subsection{Fisher forecast from topological metrics with SKA1-Low noise}
\label{subsec:fisher_noise}

To quantify how well the topological descriptors can jointly constrain X-ray heating and warm dark
matter in a realistic experiment, we perform a two-parameter Fisher-matrix forecast based on the
metrics introduced above. We take the parameter vector $\boldsymbol{\theta}=(f_X, m_{\rm WDM})$,
where $f_X$ rescales the X-ray emissivity and $m_{\rm WDM}$ is the thermal relic mass of the warm
dark matter model. For each point on the simulation grid in $(f_X,m_{\rm WDM})$, we analyze $10$
independent realizations, each providing $10$ lines of sight (LOS). The three summary statistics
$\boldsymbol{\mu}=\{\lambda(t_\star), M_2, A_{\rm skew}\}$ are measured for each LOS and then averaged
over the $10$ LOS within each realization to obtain realization-level means. The
realization-to-realization scatter $\sigma_{\rm real}$ is estimated from the dispersion of these
realization-level means across the $10$ realizations.
Because $\sigma_{\rm real}$ is estimated from the realization-to-realization scatter of the LOS-averaged metrics,
it represents an effective variance that includes finite-volume / finite-LOS sample (cosmic) variance
(in addition to the thermal-noise fluctuations propagated through the analysis pipeline). The spectra used here include SKA1-Low-like thermal noise corresponding to an integration time of $1000$~h per background source, evaluated at our fiducial spectral resolution $\Delta\nu=1$~kHz.

Assuming $N_{\rm los}$ independent background radio sources, we approximate the covariance of the
measured metrics as diagonal,
\begin{equation}
C_{\alpha\beta} \;=\; \delta_{\alpha\beta}\,
\frac{\sigma_{\rm real}^2}{N_{\rm los}},
\label{eq:cov}
\end{equation}
and neglect correlations between different metrics.
When scaling to $N_{\rm los}$ sources in Eq.~\ref{eq:cov}, we assume different sightlines are independent.
Relaxing this assumption (and including inter-metric correlations) requires a larger ensemble and a full covariance estimation,
which we leave for future work.
In principle, the three summaries are derived from the same persistence information and may
therefore exhibit non-zero cross-correlations; however, in this work we do not attempt to estimate a
full covariance matrix from our finite ensemble, and we adopt the diagonal approximation as a
simplifying assumption to illustrate the relative constraining power and degeneracy directions.

The Fisher matrix is then
\begin{equation}
F_{ij} \;=\; \sum_{\alpha,\beta}
\frac{\partial \mu_\alpha}{\partial \theta_i}
\,(C^{-1})_{\alpha\beta}\,
\frac{\partial \mu_\beta}{\partial \theta_j},
\label{eq:fisher_def}
\end{equation}
evaluated at a fiducial model $\boldsymbol{\theta}_{\rm fid}$. Throughout this section we adopt as
fiducial model a moderately heated WDM scenario with $m_{\rm WDM}=4~{\rm keV}$ and $f_X=0.2$.

\begin{figure*}[t]
  \centering
  \includegraphics[width=\hsize]{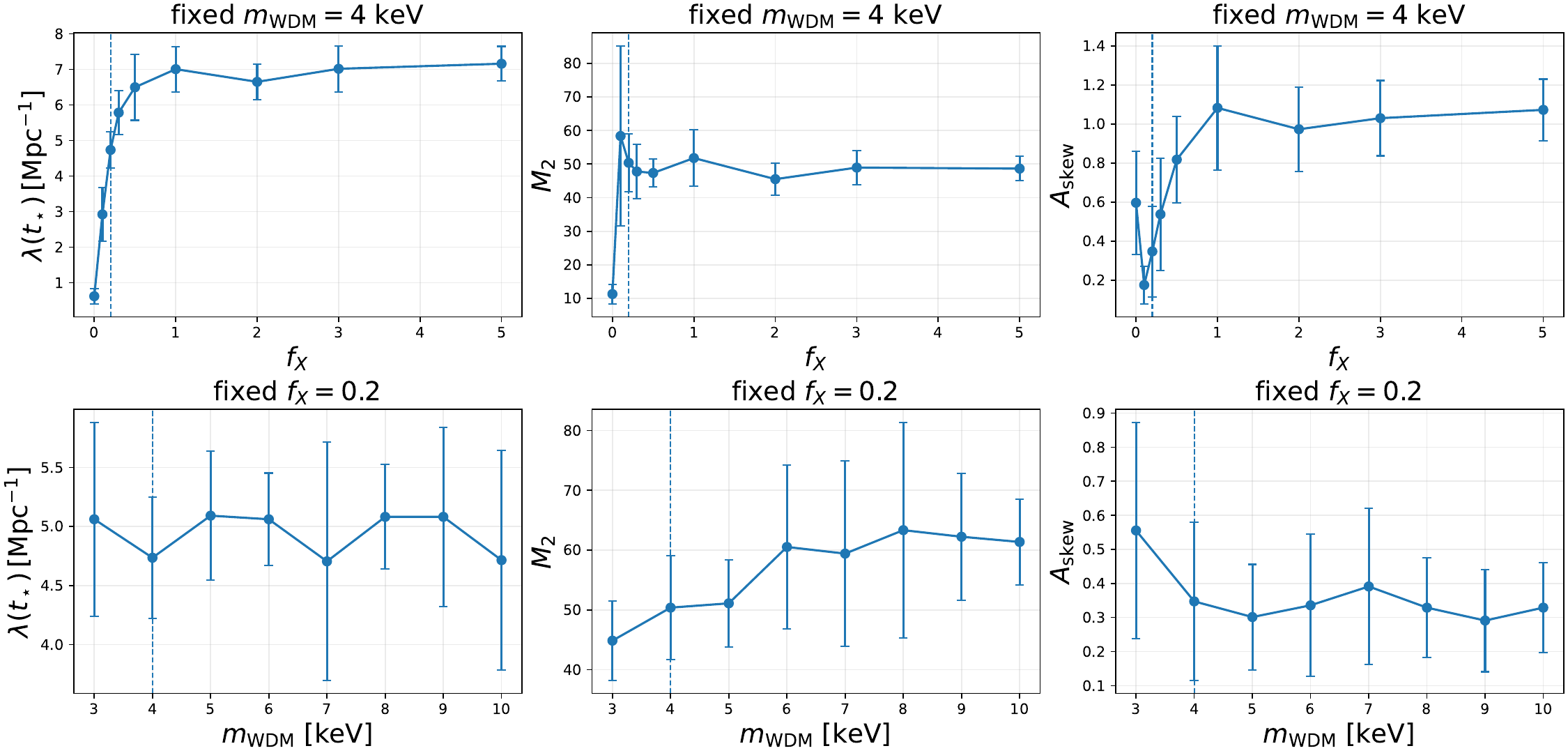}
  \caption{
  Slice plots through the fiducial model for the 1000~h noise branch.
  Top row: metric versus $f_X$ at fixed $m_{\rm WDM}=4~{\rm keV}$.
  Bottom row: metric versus $m_{\rm WDM}$ at fixed $f_X=0.2$.
  The dashed line marks the fiducial value on each axis.
  These trends explain the orientations of the single-metric degeneracy directions in Figure~\ref{fig:Fisher}: $\lambda(t_\star)$ and $A_{\rm skew}$ are locally most sensitive to $f_X$, while $M_2$ retains sensitivity to both $f_X$ and $m_{\rm WDM}$.
  }
  \label{fig:slice}
\end{figure*}

Before presenting the Fisher ellipse, it is instructive to visualize the local parameter response
of each descriptor using one-dimensional slices through the fiducial point. Figure~\ref{fig:slice}
shows $\lambda(t_\star)$, $M_2$, and $A_{\rm skew}$ as functions of $f_X$ at fixed
$m_{\rm WDM}=4~{\rm keV}$ (top row) and as functions of $m_{\rm WDM}$ at fixed $f_X=0.2$ (bottom
row), including the realization-level scatter. In the vicinity of the fiducial point,
$\lambda(t_\star)$ and $A_{\rm skew}$ exhibit a clearly stronger dependence on $f_X$ than on
$m_{\rm WDM}$ over our sampled grid. At the same time, $M_2$ retains sensitivity to both parameters
and shows a comparatively clearer trend with $m_{\rm WDM}$ in the fiducial neighborhood. This
separation of local response directions anticipates the complementarity seen in the Fisher
constraints.

\begin{figure}[htbp]
  \centering
  \includegraphics[width=\hsize]{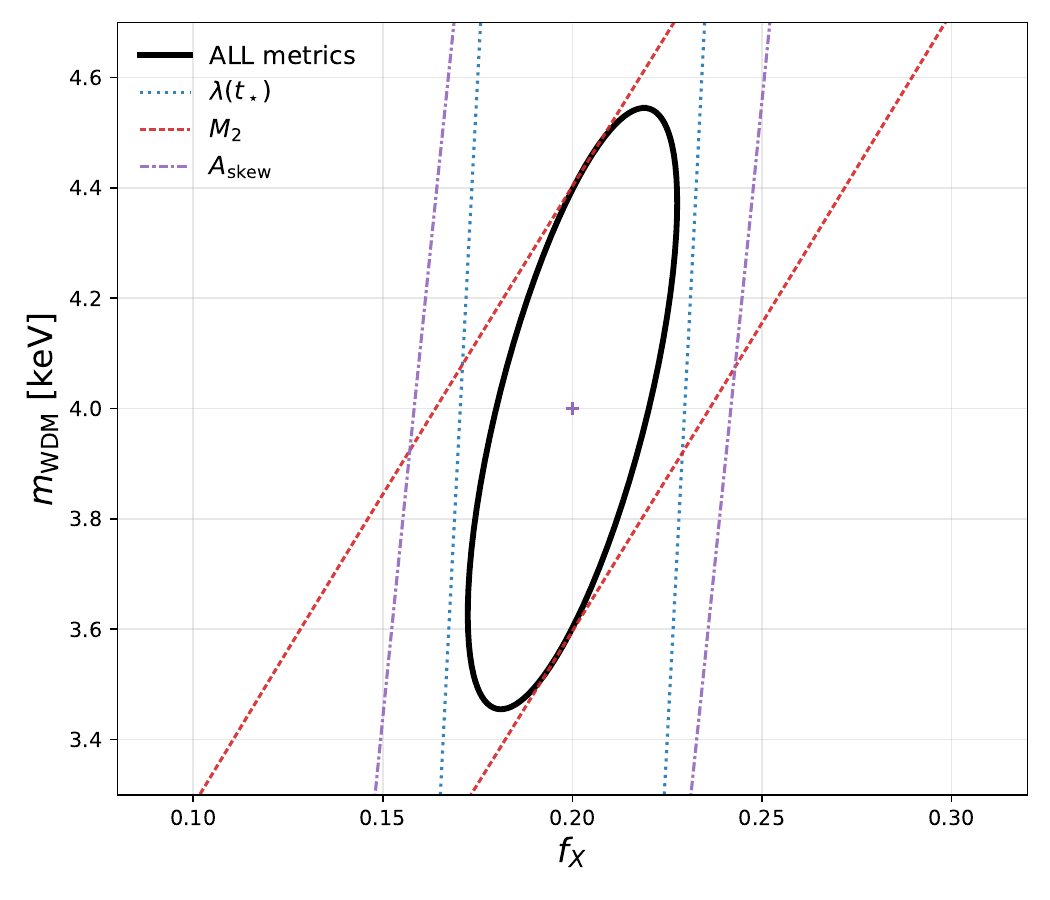}
  \caption{
  Fisher-matrix forecast for the joint constraints on $f_X$ and $m_{\rm WDM}$ using SKA1-Low--like thermal noise with 1000~h integration time assuming $N_{\rm los}$ = 10 independent background sources.
  The black contour shows the combined $1\sigma$ constraint from $\{\lambda(t_\star), M_2, A_{\rm skew}\}$ evaluated at the fiducial point $(f_X, m_{\rm WDM})=(0.2, 4~{\rm keV})$ (cross).
  Colored contours show the single-metric constraints, which are highly elongated because a single observable yields a Fisher matrix close to rank one in a two-parameter space.
  }
  \label{fig:Fisher}
\end{figure}

Figure~\ref{fig:Fisher} shows the resulting $1\sigma$ confidence region in the $(f_X,m_{\rm WDM})$
plane. The black ellipse corresponds to the combined Fisher matrix using all three statistics,
$\{\lambda(t_\star), M_2, A_{\rm skew}\}$, while the colored curves indicate the constraints obtained
when each metric is used in isolation (some are truncated by the plotting window).
The Fisher constraints shown in Fig.~\ref{fig:Fisher} are evaluated at our fiducial spectral resolution
($\Delta\nu=1$~kHz) under the thermal-noise level adopted in this section (corresponding to $t_{\rm int}=1000$~h per source).
If $\Delta\nu$ were changed, the forecast would primarily be affected through the local derivatives
$\partial\mu_\alpha/\partial\theta_i$: spectral averaging merges nearby features and reshapes persistence lifetimes,
which is expected to reduce sensitivity to the smallest-scale structure as $\Delta\nu$ is coarsened.
In a fully self-consistent observing model, increasing $\Delta\nu$ also lowers the thermal-noise amplitude via the radiometer equation
($\sigma_S\propto(\Delta\nu)^{-1/2}$), partially offsetting this loss; exploring this coupled trade-off is left for future work.

The orientations of the Fisher ellipses can be understood from the local parameter responses shown in Fig.~\ref{fig:slice}.
The orientation of each single-metric constraint is set by the local gradient of that metric at the fiducial point:
in a two-parameter Fisher forecast, the long axis points along the direction in parameter space to which the metric is least sensitive
(moving along this axis produces comparatively little change in the metric and therefore yields weaker constraints),
while the short axis aligns with the direction of maximal local sensitivity.
Strictly speaking, for a single scalar metric and no external priors the Fisher matrix is rank-1, so one direction is formally unconstrained;
in practice we therefore show highly elongated contours that may be truncated by the plotting window.
The combined ellipse reflects the eigen-directions of the total Fisher matrix, i.e., the combined gradient information from all metrics.

Consistent with Fig.~\ref{fig:slice}, the $\lambda(t_\star)$ constraint is highly elongated with a degeneracy direction close to vertical in this fiducial region.
This reflects the fact that, once thermal noise and persistence filtering are included, $\lambda(t_\star)$ varies more strongly with $f_X$ than with $m_{\rm WDM}$ near $\boldsymbol{\theta}_{\rm fid}$.
Physically, in our noise-realistic setting the component count evaluated at the reference threshold is governed primarily by heating-induced remapping and contrast compression around $t_\star$,
while the residual dependence on free streaming is subdominant at the fiducial point.

The $M_2$ constraint is tilted with an intermediate degeneracy direction.
Since $M_2$ weights long-lived troughs by $\tau^2$, it responds both to heating, which compresses contrast and shortens persistence,
and to warm-dark-matter free streaming, which modifies the abundance and lifetimes of small-scale absorbers.
This mixed sensitivity yields a diagonal response in $(f_X,m_{\rm WDM})$ and provides leverage complementary to $\lambda(t_\star)$ and $A_{\rm skew}$.

Finally, $A_{\rm skew}$ shows a sensitivity direction that is not identical to either $\lambda(t_\star)$ or $M_2$.
Because $A_{\rm skew}$ is controlled by the relative support of $\beta_0(t)$ on the deep and shallow sides, it is especially responsive to heating-driven shifts of the Betti support along the threshold axis.
Over the parameter range sampled here, its dependence on $m_{\rm WDM}$ remains comparatively weaker, so its degeneracy direction is closer to that of $\lambda(t_\star)$,
but the differing functional dependence on the full $\beta_0(t)$ shape leads to a distinct tilt in the fiducial neighborhood.

Taken together, the three descriptors probe distinct combinations of $f_X$ and $m_{\rm WDM}$ under realistic SKA1-Low noise.
The joint $1\sigma$ ellipse is therefore substantially smaller than the broad degeneracy regions from any single metric:
$\lambda(t_\star)$ and $A_{\rm skew}$ primarily restrict the heating direction in the fiducial region,
while $M_2$ supplies additional leverage along a mixed heating and free-streaming direction and reduces the remaining allowed parameter volume.
Although simplified, this forecast demonstrates that topological summaries of the 21\,cm forest remain informative for joint inference of heating and warm-dark-matter physics in the presence of thermal noise.

\

\section{Summary \& Discussion}
The topology of the 21\,cm forest provides complementary information about
the thermal and structural state of the intergalactic medium that is not
captured by traditional amplitude-based statistics.
The three topological descriptors introduced in this work—
the trough line density $\lambda(t_\star)$, the total squared persistence
$M_2$, and the Betti-curve asymmetry $A_{\rm skew}$—
probe distinct physical aspects of absorption morphology and therefore
respond to heating and dark-matter free-streaming in fundamentally
different ways.

\subsection{Persistence-based topology as a non-Gaussian probe}

It is also interesting to connect our persistence-based approach to other
non-Gaussian statistics that have been widely used for the 21\,cm
brightness-temperature field. In particular, the bispectrum (i.e., the
Fourier-space three-point function) is sensitive to phase correlations and
nonlinear mode coupling, while Minkowski functionals quantify excursion-set
morphology in real space\citep[e.g.][]{2017MNRAS.472.2436W,2018MNRAS.476.4007M,2016MNRAS.458.3003S,2017MNRAS.468.1542S,2017MNRAS.465..394Y,2018MNRAS.477.1984B,2018JCAP...10..011K,2019ApJ...885...23C,2019JCAP...09..053K,2021JCAP...05..026K,2021MNRAS.505.1863G,2021MNRAS.505.3492S}.
These ideas can be adapted to the 21\,cm forest as well, but with a
practically important caveat: forest observables are line-like absorption
features on top of a source-dependent continuum, so preprocessing choices
(e.g., detrending, windowing, masking, and normalization) can more directly
impact bispectrum- or amplitude-based morphology measures.
In this respect, persistence summaries provide a complementary viewpoint
because they emphasize the birth--merger hierarchy of troughs under sublevel
filtrations and naturally down-weight short-lived, noise-induced structures
through persistence filtering. A systematic comparison, and ultimately a
joint analysis, of persistence descriptors with bispectrum/Minkowski-type
summaries may therefore help clarify which aspects of non-Gaussianity are
uniquely captured by topology and how robust each family of statistics is
under realistic bandpass and continuum uncertainties.

\subsection{Physical interpretation of the three descriptors}

The line density $\lambda(t_\star)$ traces the abundance of independent
absorption troughs above a fixed persistence cut.
In principle, warm dark matter erases small-scale density fluctuations and
removes shallow extrema, which can reduce the number of identifiable
troughs. However, once realistic SKA1-Low thermal noise is included and
short-lived features are filtered out, the dominant variation of
$\lambda(t_\star)$ in the fiducial neighborhood is driven by the heating
parameter. In particular, around $(f_X,m_{\rm WDM})=(0.2,4~{\rm keV})$ the
slice trends show a strong change of $\lambda(t_\star)$ with $f_X$ but a
comparatively mild dependence on $m_{\rm WDM}$, implying that
$\lambda(t_\star)$ primarily constrains $f_X$ locally and yields a nearly
vertical single-metric degeneracy direction in the Fisher plane (see
Figs.~\ref{fig:slice} and \ref{fig:Fisher}). This emphasizes that Fisher
orientations are controlled by the \emph{local} gradients at the fiducial
point rather than by global monotonic trends across the full grid.

The statistic $M_2$ measures the cumulative strength of long-lived
absorption features through the total squared persistence,
$M_2=\sum_{j\in \mathcal{I}_{\mathrm{long}}}\tau_j^2$.
Both warm dark matter and X-ray heating can reduce $M_2$, but through
qualitatively different modifications of the persistence hierarchy.
Warm dark matter suppresses small-scale structure, removes shallow minima,
and accelerates mergers between neighboring troughs, thereby shifting the
lifetime distribution toward smaller $\tau$ and depleting its long-$\tau$
tail. X-ray heating, in contrast, primarily compresses the contrast of
surviving features, shortening their persistence without necessarily
eliminating the underlying rank ordering.
Because $M_2$ weights lifetimes quadratically, it is particularly sensitive
to the suppression of the long-$\tau$ tail and therefore retains
appreciable sensitivity to $m_{\rm WDM}$ even in the presence of noise and
persistence filtering. As a result, $M_2$ typically follows a moderately
inclined degeneracy direction, placing it between $\lambda(t_\star)$ and
$A_{\rm skew}$ in terms of parameter response.

The Betti-curve asymmetry $A_{\rm skew}$ encodes the relative support of
$\beta_0(t)$ on the deep and shallow sides of the distribution.
In our models, X-ray heating distorts this balance by shifting absorption
toward shallower thresholds and broadening the shallow wing of $\beta_0(t)$.
Warm dark matter primarily changes the abundance of small-scale structure
and the overall number of components, while producing a weaker change in
the deep--shallow imbalance of the surviving Betti support.
As a result, $A_{\rm skew}$ is primarily sensitive to the heating
efficiency $f_X$ and only weakly dependent on $m_{\rm WDM}$ over the
parameter range explored here, yielding a degeneracy direction that is
nearly vertical.

\subsection{Degeneracy breaking in the Fisher forecast}

Because $\lambda(t_\star)$ and $A_{\rm skew}$ both carry strong local
sensitivity to $f_X$ in the fiducial neighborhood, degeneracy breaking does
not arise simply from an orthogonality between these two metrics alone.
Instead, the key complementarity is that $M_2$ retains appreciable
sensitivity to $m_{\rm WDM}$ (and to $f_X$), supplying the inclined
direction needed to reduce the remaining degeneracy when the three metrics
are combined. In other words, in the noise-included forecast relevant for
SKA1-Low, the role of the three metrics is naturally interpreted as:
$\lambda(t_\star)$ and $A_{\rm skew}$ primarily anchor the heating axis,
while $M_2$ provides the additional leverage on the free-streaming scale
needed for a joint constraint.

We also note that the dominant topological transition associated with
heating often occurs between the strictly unheated case ($f_X=0$) and models
with $f_X>0$, while further increases in $f_X$ produce more incremental
changes. This threshold-like behavior is consistent with the rapid contrast
compression once heating is present, and it explains why the local Fisher
sensitivity to $f_X$ can be large even when global trends appear to
saturate at high heating.

\subsection{Local versus global trends}

When combined in a Fisher analysis, the complementary sensitivity directions
of $\{\lambda(t_\star),M_2,A_{\rm skew}\}$ yield substantially tighter
constraints on $(f_X,m_{\rm WDM})$ than any individual statistic alone.
The black confidence ellipse in Fig.~\ref{fig:Fisher} illustrates this
effect. It is important, however, to distinguish between (i) a statistic
showing a large dynamic range across the full $(f_X,m_{\rm WDM})$ grid and
(ii) its ability to break parameter degeneracies in a local Gaussian
forecast. The Fisher orientation and constraining power are controlled by
the local gradients evaluated at the fiducial point, not by global
monotonic trends over the entire plane. In the noise-included case this
distinction becomes especially relevant: thermal noise introduces many
short-lived, shallow fluctuations that are removed by the persistence cut,
so the remaining long-lived topology is governed by a restricted subset of
structures whose local response can differ from the global trend.
Consequently, a descriptor may vary strongly across the plane yet still
produce a highly elongated single-metric ellipse if its local response
aligns with a single direction in $(f_X,m_{\rm WDM})$.

In our forecast near $(f_X,m_{\rm WDM})=(0.2,4~{\rm keV})$,
$\lambda(t_\star)$ and $A_{\rm skew}$ provide strong local leverage on the
heating axis, while $M_2$ retains appreciable sensitivity to $m_{\rm WDM}$
(and to $f_X$), supplying an inclined constraint direction with
non-negligible free-streaming dependence. Their combination therefore
reduces the allowed parameter volume substantially compared to any
individual statistic, demonstrating that persistence-based descriptors of
the 21\,cm forest contain information that is complementary to and
qualitatively distinct from amplitude-based summaries.

\subsection{Robustness to thermal noise}

Thermal noise predominantly introduces short-lived fluctuations, which are
removed by applying a uniform persistence cut and therefore do not dominate
the long-lived components that determine $\lambda(t_\star)$, $M_2$, and
$A_{\rm skew}$. As a result, the qualitative slice trends and the principal
Fisher degeneracy directions remain stable under the uncorrelated
thermal-noise model considered here, although the overall constraining
power depends on the available number of independent background sources and
on the noise level.

\subsection{Caveats and future extensions}

Several observational and physical effects not fully explored here will be
important when applying these descriptors to real data, and we briefly
comment on their expected impact.

First, while $\sim$20 radio-loud sources may exist at $z\simeq 9$, an
initial detection of the 21\,cm forest will likely rely on a single
spectrum. In this regime, a dominant limitation can be sample variance
rather than thermal noise once a sufficient S/N is reached:
$\lambda(t_\star)$, $M_2$, and $A_{\rm skew}$ fluctuate across lines of
sight because the number and clustering of absorption troughs are set by
rare overdense structures. Our Fisher forecast assumes $N_{\rm los}$
independent background sources; in the idealized limit of independent
sightlines, the measurement uncertainties of the summary statistics
decrease approximately as $1/\sqrt{N_{\rm los}}$, and the resulting
parameter constraints typically improve in a similar manner.
For $N_{\rm los}=1$, meaningful constraints may still be achievable in
favorable cases, such as exceptionally bright background sources observed
with long integration times, but the inference becomes sensitive to cosmic
variance and prior assumptions. For early data, a practical use of the
topological descriptors is therefore model discrimination (e.g., excluding
extreme heating scenarios or very warm dark-matter models) rather than
precision parameter estimation.
We also emphasize that our Fisher estimate adopts a diagonal covariance
between the three metrics; estimating the full cross-covariance will
require substantially larger simulation ensembles and is left for future
work. Finally, quantifying the stability of the Betti curves and
persistence-based descriptors under LOS resampling (e.g., bootstrap/jackknife)
will require larger simulated ensembles and is left for future work.

Second, the present analysis targets Cosmic Dawn, where the forest forms in
a largely neutral intergalactic medium and its morphology is shaped
primarily by heating and small-scale structure. At lower redshifts
($z\lesssim 7$), where radio-loud quasars have already been identified,
inhomogeneous reionization introduces ionized bubbles and partially ionized
regions that truncate absorption segments and modify the merger hierarchy
encoded in $\beta_0(t)$. In this regime, the descriptors may become
simultaneously sensitive to ionization topology and temperature, and an
interpretation in terms of $(f_X,m_{\rm WDM})$ alone will require (at least)
additional parameters or external constraints, such as independent
estimates of the global neutral fraction or characteristic bubble sizes.
Nevertheless, the topological framework itself remains applicable: ionized
gaps and bubble boundaries generate connectivity-driven signatures that can
be captured by the same persistence formalism, suggesting a natural
extension to joint inference of heating, small-scale structure, and
reionization morphology.

Third, peculiar velocities enter the optical depth through the line-of-sight
velocity-gradient term and can both enhance absorption amplitudes and alter
the apparent clustering of features in frequency space. Because our
analysis standardizes each spectrum and emphasizes rank ordering and
connectivity, purely multiplicative amplitude boosts are expected to have a
limited impact. However, velocity-induced reordering of shallow extrema or
shifts in the relative separations of neighboring troughs can affect
$\lambda(t_\star)$ and the population of short-lived components.
A quantitative assessment including full redshift-space distortions is
therefore an important next step, particularly to test whether the nearly
orthogonal responses of $\{\lambda(t_\star),M_2,A_{\rm skew}\}$ persist when velocity effects are modeled self-consistently. 

Fourth, another physical effect not explicitly modeled here is the quasar proximity (near-zone) effect. 
A luminous background quasar can ionize and heat the proximate IGM, producing a local suppression of 21-cm absorption near the systemic redshift (e.g., \citep{Soltinsky2023}). 
If this region is included in the analysis window, the reduction of absorption and the modification of trough structure could bias the topological descriptors, for example by lowering the effective trough abundance or shortening the persistence of features, partly mimicking stronger heating. 
In practice, however, the proximity effect is localized in frequency (or comoving distance) and can be mitigated by masking a window around the systemic redshift prior to detrending and topological analysis. 
Alternatively, simple nuisance parameters describing the near-zone extent can be incorporated into forward models and marginalized over in inference. 
A detailed radiative-transfer treatment of the quasar near-zone is left for future work.

Finally, in realistic observations the thermal-noise level depends on the
flux density and intrinsic spectral shape of the background radio source,
as well as on the channel width $\Delta\nu$ through the radiometer
equation. In this work, spectral resolution and integration time were
varied separately to isolate their qualitative effects on topology. A fully
consistent forward model would vary the noise amplitude together with
$\Delta\nu$ and marginalize over uncertainties in the background continuum,
calibration residuals, and bandpass structure, as well as possible
frequency correlations in the noise. While such refinements can shift the
absolute performance forecasts, the central trends reported here are driven
by long-lived troughs and the gross shape of $\beta_0(t)$, which are
comparatively stable once short-lived components are removed.

Because noise-induced features are preferentially short-lived, persistence
filtering provides a natural mitigation. In practice, the optimal choice
of $\tau_\star$ and the associated information loss should be calibrated
under realistic observing conditions; however, varying $\tau_\star$ within
a reasonable range does not change the qualitative parameter responses of
$\{\lambda(t_\star), M_2, A_{\rm skew}\}$ in our simulations.

\section{CONCLUSION}

We have shown that persistence-based topology provides a physically
interpretable and complementary analysis layer for the 21\,cm forest,
helping to disentangle X-ray heating from dark-matter free-streaming.
By tracking the birth--merger hierarchy of absorption troughs through
sublevel filtrations of detrended, standardized one-dimensional forest
spectra, we construct persistence diagrams and Betti--0 curves that encode
connectivity and ordering information beyond two-point statistics.

From these topological summaries we defined three descriptors with clear
interpretations: the trough line density $\lambda(t_\star)$, the total
squared persistence $M_2$, and the Betti-curve asymmetry $A_{\rm skew}$.
In our SKA1-Low noise-included forecasts, $\lambda(t_\star)$ and
$A_{\rm skew}$ provide strong local leverage on the heating axis, while
$M_2$ retains appreciable sensitivity to the free-streaming scale and
supplies the additional constraint direction needed to reduce the remaining
degeneracy. More generally, the relative sensitivity of each descriptor can
vary across the parameter plane because Fisher information is governed by
local gradients; the key result is that the three descriptors probe
distinct aspects of the merger hierarchy and Betti support and therefore
provide non-redundant information when combined.

The same persistence-based framework is readily applicable to other
one-dimensional absorption fields such as the Ly$\alpha$ forest, offering a
unified route to non-Gaussian inference that emphasizes connectivity and
merger hierarchy. Applied to high-quality data from next-generation
facilities such as the SKA, these topology-based descriptors provide a
geometry-aware window on early structure formation and the microphysics of
dark matter, naturally complementing power-spectrum and wavelet-based
analyses.

\begin{acknowledgments}
This work is supported by the National SKA Program of China (No.2020SKA0110401), NSFC (Grant No.~12103044), and Yunnan Provincial Key Laboratory of Survey Science with project No. 202449CE340002. I'm grateful for the referee’s careful reading and insightful suggestions, which helped me sharpen the clarity and scope of this work.
\end{acknowledgments}

\bibliography{chapter}

\end{document}